\documentclass{article}
\usepackage{ismir,amsmath,cite,url}
\usepackage{amssymb}
\usepackage{graphicx}
\usepackage{color}

\usepackage{pgfplots}
\usetikzlibrary{backgrounds}

\usepackage[utf8]{inputenc}

\usepackage[colorinlistoftodos]{todonotes}
\usepackage{bm}

\newcommand{\quotes}[1]{``#1''}

\DeclareMathOperator{\lagr}{\mathcal{L}}

\usepackage{makecell}
\usepackage{array}
\newcolumntype{?}{!{\vrule width 1pt}}

\newcommand{\ts}{\textsuperscript}

\title{MIDI-VAE: Modeling Dynamics and Instrumentation of Music with Applications to Style Transfer}

\multauthor
{Gino Brunner \hspace{1cm} Andres Konrad \hspace{1cm} Yuyi Wang\hspace{1cm} Roger Wattenhofer} {
Department of Electrical Engineering and Information Technology\\ ETH Zurich\\ Switzerland\\
{\tt\small brunnegi,konradan,yuwang,wattenhofer@ethz.ch }
}
\def\authorname{Gino Brunner, Andres Konrad, Yuyi Wang, Roger Wattenhofer}

\begin{document}

\maketitle
\begin{abstract}
We introduce MIDI-VAE, a neural network model based on Variational Autoencoders that is capable of handling polyphonic music with multiple instrument tracks, as well as modeling the dynamics of music by incorporating note durations and velocities. We show that MIDI-VAE can perform style transfer on symbolic music by automatically changing pitches, dynamics and instruments of a music piece from, e.g., a Classical to a Jazz style. We evaluate the efficacy of the style transfer by training separate style validation classifiers. Our model can also interpolate between short pieces of music, produce medleys and create mixtures of entire songs. The interpolations smoothly change pitches, dynamics and instrumentation to create a harmonic bridge between two music pieces.  
To the best of our knowledge, this work represents the first successful attempt at applying neural style transfer to complete musical compositions. 
\end{abstract}
\section{Introduction}\label{sec:introduction}

Deep generative models do not just allow us to generate new data, but also to change properties of existing data in principled ways, and even transfer properties between data samples. 
Have you ever wanted to be able to create paintings like Van Gogh or Monet? No problem! Just take a picture with your phone, run it through a neural network, and out comes your personal masterpiece.
Being able to generate new data samples and perform style transfer requires models to obtain a deep understanding of the data. Thus, advancing the state-of-the-art in deep generative models and neural style transfer is not just important for transforming horses into zebras,\footnote{\url{https://junyanz.github.io/CycleGAN/}} but lies at the very core of Deep (Representation) Learning research~\cite{DBLP:journals/pami/BengioCV13}. 

While neural style transfer has produced astonishing results especially in the visual domain~\cite{DBLP:conf/nips/LiFYWLY17,DBLP:conf/iccv/ZhuPIE17}, the progress for sequential data, and in particular music, has been slower. We can already transfer sentiment between restaurant reviews~\cite{DBLP:conf/nips/ShenLBJ17,DBLP:journals/corr/ZhaoKZRL17}, or even change the instrument with which a melody is played~\cite{DBLP:conf/nips/OordVK17}, but we have no way of knowing how our favorite pop song would have sounded if it were written by a composer who lived in the classical epoch or how a group of jazz musicians would play the Overture of Mozart's Don Giovanni.
In this work we take a step towards this ambitious goal. To the best of our knowledge, this paper presents the first successful application of unaligned style transfer to musical compositions. Our proposed model architecture consists of parallel Variational Autoencoders (VAE) with a shared latent space and an additional style classifier. The style classifier forces the model to encode style information in the shared latent space, which then allows us to manipulate existing songs, and effectively change their style, e.g., from Classic to Jazz. Our model is capable of producing harmonic polyphonic music with multiple instruments. It also learns the dynamics of music by incorporating note durations and velocities.

\section{Related Work}\label{sec:relwork}



Gatys et al.~\cite{DBLP:conf/cvpr/GatysEB16} introduce the concept of neural style transfer and show that pre-trained CNNs can be used to merge the style and content of two images. Since then, more powerful approaches have been developed \cite{DBLP:conf/nips/LiFYWLY17,DBLP:conf/iccv/ZhuPIE17}; these allow, for example, to render an image taken in summer to look like it was shot in winter. 
For sequential data, autoencoder based methods \cite{DBLP:conf/nips/ShenLBJ17,DBLP:journals/corr/ZhaoKZRL17} have been proposed to change the sentiment or content of sentences. 
Van den Oord et al.~\cite{DBLP:conf/nips/OordVK17} introduce a VAE model with discrete latent space that is able to perform speaker voice transfer on raw audio data. Mor et al.~\cite{DBLP:journals/corr/abs-1805-07848} develop a system based on a WaveNet autoencoder~\cite{DBLP:conf/icml/EngelRRDNES17} that can translate music across instruments, genres and styles, and even create music from whistling. 
Malik et al.~\cite{DBLP:journals/corr/abs-1708-03535} train a model to add note velocities (loudness) to sheet music, resulting in more realistic sounding playback. Their model is trained in a supervised manner, with the target being a human-like performance of a music piece in MIDI format, and the input being the same piece but with all note velocities set to the same value. While their model can indeed 
play music in a more human-like manner, it can only change note velocities, and does not learn the characteristics of different musical styles/genres.
Our model is trained on unaligned songs from different musical styles. Our model can not only change the dynamics of a music piece from one style to another, but also automatically adapt the instrumentation and even the note pitches themselves. 
Apart from style transfer, our model can also generate short pieces of music, medleys, interpolations and song mixtures. At the core of our model thus lies the capability to produce music. In the following we will therefore discuss related work in the domains of symbolic and raw audio generation.
For a more comprehensive overview we refer the interested readers to these surveys: \cite{DBLP:journals/jair/FernandezV13,briot2017deep,DBLP:journals/csur/HerremansCC17}.

\begin{figure*}[ht]
\centering
\includegraphics[width=\textwidth]{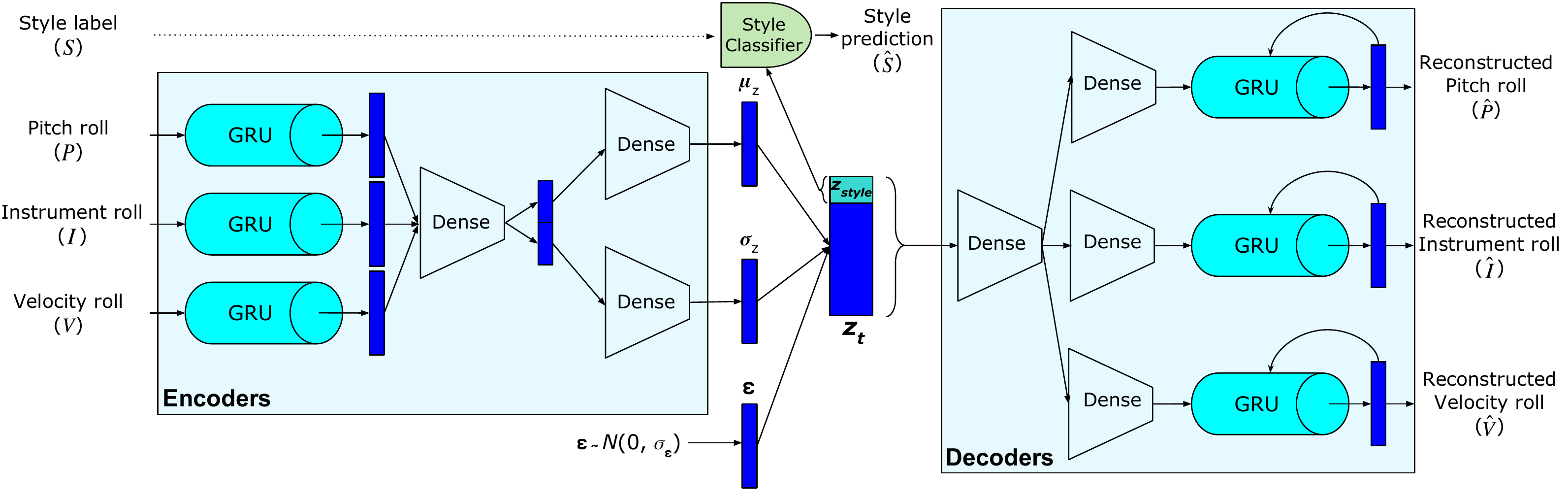}
\caption{
MIDI-VAE architecture. GRU stands for Gated Recurrent Unit~\cite{DBLP:conf/emnlp/ChoMGBBSB14}.}
\label{fig:modelarch}
\end{figure*}

People have been trying to compose music with the help of computers for decades. One of the most famous early examples is \quotes{Experiments in Musical Intelligence}~\cite{DBLP:conf/icmc/Cope87}, a semi-automatic system based on Markov models that is able to create music in the style of a certain composer. 
Soon after, the first attempts at music composition with artificial neural networks were made. Most notably, Todd~\cite{todd1989connectionist}, Mozer~\cite{DBLP:journals/connection/Mozer94} and Eck et al.~\cite{eck2002first} all used Recurrent Neural Networks (RNN).
More recently, Boulanger-Lewandowski et al.~\cite{DBLP:conf/icml/Boulanger-LewandowskiBV12} combined long short term memory networks (LSTMs) and Restricted Boltzmann Machines to simultaneously model the temporal structure of music, as well as the harmony between notes that are played at the same time, thus being capable of generating polyphonic music.
Chu et al.~\cite{DBLP:journals/corr/ChuUF16} use domain knowledge to model a hierarchical RNN architecture that produces multi-track polyphonic music. 
Brunner et al.~\cite{brunner2017jambot} combine a hierarchical LSTM model with learned chord embeddings that form the Circle of Fifths, showing that even simple LSTMs are capable of learning music theory concepts from data. 
Hadjeres et al.~\cite{DBLP:conf/icml/HadjeresPN17} introduce an LSTM-based system that can harmonize melodies by composing accompanying voices in the style of Bach Chorales, which is considered a very difficult task even for professionals. 
Johnson et al.~\cite{DBLP:conf/evoW/Johnson17} use parallel LSTMs with shared weights to achieve transposition-invariance (similar to the translation-invariance of CNNs). 
Chuan et al.~\cite{chuan2018modeling} investigate the use of an image-based Tonnetz representation of music, and apply a hybrid LSTM/CNN model to music generation.

Generative models such as the Variational Autoencoder (VAE) and Generative Adversarial Networks (GANs) have been increasingly successful at modeling music.
Roberts et al. introduce MusicVAE~\cite{roberts2017hierarchical}, a hierarchical VAE model that can capture long-term structure in polyphonic music and exhibits high interpolation and reconstruction performance. 
GANs, while very powerful, are notoriously difficult to train and have generally not been applied to sequential data. However, Mogren~\cite{DBLP:journals/corr/Mogren16}, Yang et al.~\cite{DBLP:conf/ismir/YangCY17} and Dong et al.~\cite{DBLP:journals/corr/abs-1709-06298} have recently shown the efficacy of CNN-based GANs for music composition.
Yu et al.~\cite{DBLP:conf/aaai/YuZWY17} were the first to successfully apply RNN-based GANs to music by incorporating reinforcement learning techniques.

Researchers have also worked on generating raw audio waves. Van den Oord et al.~\cite{DBLP:conf/ssw/OordDZSVGKSK16} introduce WaveNet, a CNN-based model for the conditional generation of speech. The authors also show that it can be used to generate pleasing sounding piano music. More recently, Engel et al.~\cite{DBLP:conf/icml/EngelRRDNES17} incorporated WaveNet into an Autoencoder structure to generate musical notes and different instrument sounds. 
Mehri et al.~\cite{DBLP:journals/corr/MehriKGKJSCB16} developed SampleRNN, an RNN-based model for unconditional generation of raw audio. While these models are impressive, the domain of raw audio is very high dimensional and it is much more difficult to generate pleasing sounding music. Thus most existing work on music generation uses symbolic music representations (see e.g., \cite{DBLP:journals/corr/abs-1708-03535,DBLP:conf/icmc/Cope87,todd1989connectionist,DBLP:journals/connection/Mozer94,DBLP:conf/icml/Boulanger-LewandowskiBV12,DBLP:journals/corr/ChuUF16,brunner2017jambot,DBLP:conf/icml/HadjeresPN17,DBLP:conf/evoW/Johnson17,chuan2018modeling,roberts2017hierarchical,DBLP:journals/corr/Mogren16,DBLP:conf/ismir/YangCY17,DBLP:journals/corr/abs-1709-06298,DBLP:conf/aaai/YuZWY17}).

\section{Model Architecture}

Our model is based on the Variational Autoencoder~\cite{DBLP:journals/corr/KingmaW13} (VAE) and operates on a symbolic music representation that is extracted from MIDI~\cite{midistandard} files. We extend the standard piano roll representation of note pitches with velocity and instrument rolls, modeling the most important information contained in MIDI files. Thus, we term our model MIDI-VAE. MIDI-VAE uses separate recurrent encoder/decoder pairs that share a latent space. A style classifier is attached to parts of the latent space to make sure the encoder learns a compact latent style label that we can then use to perform style transfer. The architecture of MIDI-VAE is shown in Figure~\ref{fig:modelarch}, and will be explained in more detail in the following.

\subsection{Symbolic Music Representation} \label{subsec:symbolicrep}

We use music files in the MIDI format, which is a symbolic representation of music that resembles sheet music. MIDI files have multiple tracks. Tracks can either be \emph{on} with a certain pitch and velocity, \emph{held} over multiple time steps or be \emph{silent}. Additionally, an instrument is assigned to each track. 
To feed the note pitches into the model we represent them as a tensor $P\in \{0,1\}^{n_P \cdot n_B \cdot n_T}$ (commonly known as piano roll and henceforth referred to as pitch roll), where $n_P$ is the number of possible pitch values, $n_B$ is the number of beats and $n_T$ is the number of tracks. 
Thus, each song in the dataset is split into pieces of length $n_B$. We choose $n_B$ such that each piece corresponds to one bar. 
We include a \quotes{silent} note pitch to indicate when no note is played at a time step. The note velocities are encoded as tensor $V\in[0,1]^{n_P \cdot n_B \cdot n_T}$ (velocity roll). 
Velocity values between 0.5 and 1 signify a note being played for the first time, whereas a value below 0.5 means that either no note is being played, or that the note from the last time step is being held. The note velocity range defined by MIDI (0 to 127) is mapped to the interval $[0.5,1]$.
We model the assignment of instruments to tracks as matrix $I=\{0,1\}^{n_T \cdot n_I}$ (instrument roll), where $n_I$ is the number of possible instruments. The instrument assignment is a global property and thus remains constant over the duration of one song. Finally, each song in our dataset belongs to a certain style, designated by the style label $S\in \{Classic, Jazz, Pop, Bach, Mozart\}$. 

In order to generate harmonic polyphonic music it is important to model the joint probability of simultaneously played notes. A standard recurrent neural network model already models the joint distribution of the sequence through time. If there are multiple outputs to be produced per time step, a common approach is to sample each output independently. In the case of polyphonic music, this can lead to dissonant and generally \quotes{wrong} sounding note combinations. However, by unrolling the piano rolls in time we can let the RNN learn the joint distribution of simultaneous notes as well. 
Basically, instead of one $n_T$-hot vector for each beat, we input $n_T$ 1-hot vectors per beat to the RNN. This is a simple but effective way of modeling the joint distribution of notes. The drawback is that the RNN needs to model longer sequences.
We use the pretty\_midi~\cite{prettymidi} Python library to extract information from MIDI files and convert them to piano rolls.

\subsection{Parallel VAE with Shared Latent Space}

MIDI-VAE is based on the standard VAE~\cite{DBLP:journals/corr/KingmaW13} with a hyperparameter $\beta$ to weigh the Kullback-Leibler divergence in the loss function (as in~\cite{higgins2016beta}). 
A VAE consists of an encoder $q_\theta(z|x)$, a decoder $p_\phi(x|z)$ and a latent variable $z$, where $q$ and $p$ are usually implemented as neural networks parameterized by $\theta$ and $\phi$. 
In addition to minimizing the standard autoencoder reconstruction loss, VAEs also impose a prior distribution $p(z)$ on the latent variables. Having a known prior distribution enables generation of new latent vectors by sampling from that distribution. Furthermore, the model will only \quotes{use} a new dimension, i.e., deviate from the prior distribution, if it significantly lowers the reconstruction error. This encourages disentanglement of latent dimensions and helps learning a compact hidden representation. The VAE loss function is 
\begin{equation*}\label{eqn:1}
\lagr_{VAE} = \mathbb{E}_{q_\theta(z|x)}[\log p_\phi(x|z)] - \beta D_{KL}[(q_\theta(z|x)||p(z)],
\end{equation*}
where the first term corresponds to the reconstruction loss, and the second term forces the distribution of latent variables to be close to a chosen prior. 
$D_{KL}$ is the Kullback-Leibler divergence, which gives a measure of how similar two probability distributions are. As is common practice, we use an isotropic Gaussian distribution with unit variance as our prior, i.e., $p(z) = \mathcal{N}(0,I)$. Thus, both $q_{\theta}(z|x)$ and $p(z)$ are (isotropic) Gaussian distributions and the KL divergence can be computed in closed form. 



As described in Section~\ref{subsec:symbolicrep}, we represent multi-track music as a combination of note pitches, note velocities and an assignment of instruments to tracks. In order to generate harmonic multi-track music, we need to model a joint distribution of these input features instead of three marginal distributions. Thus, our model consists of three encoder/decoder pairs with a shared latent space that captures the joint distribution.  
For each input sample (i.e., a piece of length $n_B$ beats), the pitch, velocity and instrument rolls are passed through their respective encoders, implemented as RNNs. The output of the three encoders is concatenated and passed through several fully connected layers, which then predict $\bm{\sigma}_z$ and $\bm{\mu}_z$, the parameters of the approximate posterior $q_\theta(z|x)=\mathcal{N}(\bm{\mu}_z,\bm{\sigma}_{z}).$\footnote{We use notation $\bm{\sigma}$ for both a variance vector and the corresponding diagonal variance matrix.} Using the reparameterization trick~\cite{DBLP:journals/corr/KingmaW13}, a latent vector $\bm{z}$ is sampled from this distribution as $\bm{z}\sim\mathcal{N}(\bm{\mu}_z,\bm{\sigma}_z*\bm{\epsilon})$ where~$*$ stands for element-wise multiplication. This is necessary because it is generally not possible to backpropagate gradients through a random sampling operation, since it is not differentiable.
$\bm{\epsilon}$ 
is sampled from an isotropic Gaussian distribution $\mathcal{N}(0,\sigma_\epsilon*I)$, where we treat $\sigma_\epsilon$ as a hyperparameter (see Section \ref{subsec:hyperparams} for more details).
This shared latent vector is then fed into three parallel fully connected layers, from which the three decoders try to reconstruct the pitch, velocity and instrument rolls. The note pitch and instrument decoders are trained with cross entropy losses, whereas for the velocity decoder we use MSE.

\subsection{Style Classifier}

Having a disentangled latent space might enable some control over the style of a song. If for example one dimension in the latent space encodes the dynamics of the music, then we could easily change an existing piece by only varying this dimension. Choosing a high value for $\beta$ (the weight of the KL term in the VAE loss function) has been shown to increase disentanglement of the latent space in the visual domain~\cite{higgins2016beta}. However, increasing $\beta$ has a negative effect on the reconstruction performance. 
Therefore, we introduce additional structure into the latent space by attaching a softmax style classifier to the top $k$ dimensions of the latent space ($z_{style}$), where $k$ equals the number of different styles in our dataset. This forces the encoder to write a \quotes{latent style label} into the latent space. Using only $k$ dimensions and a weak classifier encourages the encoder to learn a compact encoding of the style. In order to change a song's style from $S_i$ to $S_j$, we pass the song through the encoder to get $\bm{z}$, swap the values of dimensions $z_{style}^i$ and $z_{style}^j$, and pass the modified latent vector through the decoder. As style we choose the music genre (e.g., Jazz, Pop or Classic) or individual composers (Bach or Mozart).



\subsection{Full Loss Function}

Putting all parts together, we get the full loss function of our model as
\begin{align}\label{eqn:2}
\lagr_{tot} = &\lambda_P  H(P,\hat{P}) + \lambda_I H(I,\hat{I}) \\ 
+& \lambda_V MSE(V,\hat{V}) + \lambda_S H(S,\hat{S}) - \beta D_{KL}(q||p), \nonumber
\end{align}
where $H(\cdot,\cdot)$, $MSE(\cdot,\cdot)$ and $D_{KL}(\cdot||\cdot)$ stand for cross entropy, mean squared error and KL divergence respectively. The hats denote the predicted/reconstructed values. The weights $\lambda$ and $\beta$ can be used to balance the individual terms of the loss functions.

\section{Implementation}
In this section we describe our dataset and pre-processing steps. We also give some insight into the training of our model and justification for hyperparameter choices. 

\subsection{Dataset and Pre-Processing}

\begin{table}[t]
\centering
\begin{tabular}{|l?l|l|p{3.5cm}|}
\hline
Dataset & \#Songs &\#Bars & Artists \\ \Xhline{1pt}
Classic  & 477 & 60523 & Beethoven, Clementi, ... \\ \hline
Jazz & 554 & 72190 & Sinatra, Coltrane, ...  \\ \hline
Pop & 659& 65697 & ABBA, Bruno Mars, ... \\ \hline
Bach & 156& 16213 & Bach \\ \hline
Mozart & 143& 17198 & Mozart \\ \hline
\end{tabular}
\caption{Properties of our dataset.}
\label{tab:dataset}
\end{table}

Our dataset contains songs from the genres Classic, Jazz and Pop. The songs were gathered from various online sources;\footnote{Pop: \url{www.midiworld.com} / Jazz: \url{http://midkar.com/jazz/jazz_01.html} / Classic (including Bach, Mozart): \url{www.reddit.com/r/WeAreTheMusicMakers/comments/3ajwe4/}} a summary of the properties is shown in Table~\ref{tab:dataset}. Note that we excluded symphonies from our Classic, Bach and Mozart datasets due to their complexity and high number of simultaneously playing instruments. We use a train/test split of 90/10.
Each song in the dataset can contain multiple instrument tracks and each track can have multiple notes played at the same time. Unless stated otherwise, we select $n_T=4$ instrument tracks from each song by first picking the tracks with the highest number of played notes, and from each track we choose the highest voice, meaning picking the highest notes per time step. If a song has fewer than $n_T$ instrument tracks, we pick additional voices from the tracks until we have $n_T$ voices in total. We exclude drum tracks, since they do not have a pitch value.
We choose the 16\ts{th} note as smallest unit. In the most widely used time signature $\frac{4}{4}$ there are 16 16\ts{th} notes in a bar. 91\% of Jazz and Pop songs in our dataset are in $\frac{4}{4}$, whereas for Classic the fraction is 34\%. 
For songs with time signatures other than $\frac{4}{4}$ we still designate 16 16th notes as one bar. 
All songs are split into samples of one bar and our model auto-encodes one sample at a time. During training we shuffle the songs for each epoch, but keep the bars of a song in the correct order and do not reset the RNN states between samples. Thus, our model is trained on a proper sequence of bars, instead of being confused by random bar progressions. 

There are 128 possible pitches in MIDI. Since very low and high pitches are rare and often do not sound pleasing, we only use $n_P=60$ pitch values ranging from 24 ($C_1$) to 84 ($C_6$).


\subsection{Model (Hyper-)Parameters}\label{subsec:hyperparams}




Our model is generally not sensitive to most hyperparameters. Nevertheless we continuously performed local hyperparameter searches based on good baseline models, only varying one hyperparameter at a time. We use the reconstruction accuracy of the pitch roll decoder as evaluation metric. Using Gated Recurrent Units (GRUs)~\cite{DBLP:conf/emnlp/ChoMGBBSB14} instead of LSTMs increases performance significantly. Using bidirectional GRUs did not improve the results. The pitch roll encoder/decoder uses two GRU layers, whereas the rest uses only one layer. All GRU state sizes as well as the size of the latent space $z$ are set to 256. We use the ADAM optimizer~\cite{DBLP:journals/corr/KingmaB14} with an initial learning rate of 0.0002. For most layers in our architecture, we found tanh to work better than sigmoid or rectified linear units. We train on batches of size 256.  
The loss function weights $\lambda_P$, $\lambda_I$, $\lambda_V$ and $\lambda_S$ were set to 1.0, 1.0, 0.1 and 0.1 respectively. $\lambda_p$ was set to 1.0 to favor high quality note pitch reconstructions over the rest. $\lambda_V$ was also set to 1.0 because the MSE magnitude is much smaller than the cross entropy loss values. 




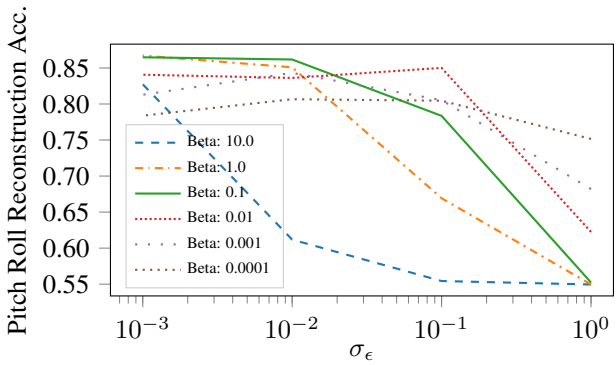
\begin{figure}[t]
\centering
\begin{tikzpicture}

\definecolor{color0}{rgb}{0.12156862745098,0.466666666666667,0.705882352941177}
\definecolor{color3}{rgb}{0.83921568627451,0.152941176470588,0.156862745098039}
\definecolor{color4}{rgb}{0.580392156862745,0.403921568627451,0.741176470588235}
\definecolor{color1}{rgb}{1,0.498039215686275,0.0549019607843137}
\definecolor{color2}{rgb}{0.172549019607843,0.627450980392157,0.172549019607843}
\definecolor{color5}{rgb}{0.549019607843137,0.337254901960784,0.294117647058824}

\begin{axis}[
xlabel={$\sigma_{\epsilon}$},
ylabel={Pitch Roll Reconstruction Acc.},
xmin=0.000607945784384139, xmax=1.41253754462275,
ymin=0.533663075897538, ymax=0.883058500064946,
height = 0.60\columnwidth,
width= 1.0\columnwidth,
xmode=log,
xtick={1e-05,0.0001,0.001,0.01,0.1,1,10,100},
xticklabels={$10^{-5}$,$10^{-4}$,$10^{-3}$,$10^{-2}$,$10^{-1}$,$10^{0}$,$10^{1}$,$10^{2}$},
ytick={0.5,0.55,0.6,0.65,0.7,0.75,0.8,0.85,0.9},
yticklabels={0.50,0.55,0.60,0.65,0.70,0.75,0.80,0.85,0.90},
tick align=outside,
tick pos=left,
x grid style={lightgray!92.02614379084967!black},
y grid style={lightgray!92.02614379084967!black},
legend cell align={left},
legend style={at={(0.03,0.03)}, anchor=south west, draw=white!80.0!black, font=\tiny, fill=none},
legend entries={{Beta: 10.0},{Beta: 1.0},{Beta: 0.1},{Beta: 0.01},{Beta: 0.001},{Beta: 0.0001}}
]
\addlegendimage{no markers, color0, dashed, thick}
\addlegendimage{no markers, color1,dashdotted, thick}
\addlegendimage{no markers, color2, thick}
\addlegendimage{no markers, color3, densely dotted, thick}
\addlegendimage{no markers, color4, loosely dotted, thick}
\addlegendimage{no markers, color5, dotted, thick}
\addplot [thick, color0, dashed]
table {%
1 0.549544686086965
0.1 0.554330248782514
0.01 0.611815060288938
0.001 0.827379658124794
};
\addplot [thick, color1, dashdotted]
table {%
1 0.549546092925563
0.1 0.669208481396159
0.01 0.850884461081067
0.001 0.867176889875518
};
\addplot [thick, color2]
table {%
1 0.55226648847399
0.1 0.783504715863829
0.01 0.861739100979037
0.001 0.864804829706785
};
\addplot [thick, color3,densely dotted]
table {%
1 0.622474812590917
0.1 0.850092921828988
0.01 0.83615603173379
0.001 0.840622539775086
};
\addplot [thick, color4, loosely dotted]
table {%
1 0.681979140055659
0.1 0.805310639710232
0.01 0.843174112788324
0.001 0.812846656528773
};
\addplot [thick, color5, dotted]
table {%
1 0.751471941225575
0.1 0.804555928977962
0.01 0.806602212593752
0.001 0.783772318700316
};
\end{axis}

\end{tikzpicture}
\caption{Test reconstruction accuracy of pitch roll for different $\beta$ and $\sigma_{\epsilon}$.}
\label{fig:betaepseval}
\end{figure}

During our experiments, we realized that high values of $\beta$ generally lead to very poor performance. 
We further found that setting the variance of $\bm{\epsilon}$ to the value of $\sigma_\epsilon=1$, as done in all previous work using VAEs, also has a negative effect. Therefore we decided to treat $\sigma_\epsilon$ as a hyperparameter as well. Figure~\ref{fig:betaepseval} shows the results of the grid search. 
$\sigma_\epsilon$ is the variance of the distribution from which the $\bm{\epsilon}$ values for the reparameterization trick are sampled, and is thus usually set to the same value as the variance of the prior. However, especially at the beginning of learning, this introduces a lot of noise that the decoder needs to handle, since the values for $\bm{\mu}_z$ and $\bm{\sigma}_z$, output by the encoder, are small compared to $\bm{\epsilon}$. 
We found that by reducing $\sigma_\epsilon$, we can improve the performance of our model significantly, while being able to use higher values for $\beta$. An annealing strategy for both $\beta$ and $\sigma_\epsilon$ might produce better results, but we did not test this. In the final models we use $\beta=0.1$ and $\sigma_\epsilon=0.01$. Note that during generation we sample $\bm{z}$ from $\mathcal{N}(0,\bm{\sigma}_{\hat{z}})$, where $\bm{\sigma}_{\hat{z}}$ is the empirical variance obtained by feeding the entire training dataset through the encoder. The empirical mean $\bm{\mu}_{\hat{z}}$ is very close to zero.



\subsection{Training}\label{subsec:training}

All models are trained on single GPUs (GTX 1080) until the pitch roll decoder converges. This corresponds to around 400 epochs, or 48 hours. 
We train one model for each genre/composer pair to make learning easier. This results in four models that we henceforth call CvJ (trained on Classic and Jazz), CvP (Classic and Pop), JvP (Jazz and Pop) and BvM (Bach and Mozart). The train/test accuracies/losses of all final models are shown in Table~\ref{tab:traintestfinal}. The columns correspond to the terms in our model's full loss function (Equation~\ref{eqn:2}).

\begin{table}[t]
\centering
\tabcolsep=0.085cm
\begin{tabular}{|l?l|l?l|l?l|l?l|l|l|}
\hline
 & \multicolumn{2}{c?}{\textbf{P}itch} & \multicolumn{2}{c?}{\textbf{I}nstrument}  & \multicolumn{2}{c?}{\textbf{S}tyle}  & \multicolumn{2}{c|}{\textbf{V}elocity} \\ \hline
  & Train & Test & Train & Test & Train & Test & Train & Test\\ \Xhline{1pt}
CvJ & 0.90 & 0.85 & 0.99 & 0.87 & 0.98&  \textbf{0.92} & 0.008 & \textbf{0.029}\\ \hline
CvP	& 0.96 & \textbf{0.88} & 0.99 & \textbf{0.89} & 0.96 & 0.91 & 0.017 & 0.036\\ \hline
JvP	& 0.88 & 0.80 & 0.99 & 0.86 & 0.94 & 0.69 & 0.043 & 0.048\\ \hline
BvM	& 0.91 & 0.75 & 0.99 & 0.82 & 0.94 & 0.74 & 0.010 & 0.033\\ \hline
\end{tabular}
\caption{Train and test performance of our final models. The velocity column shows MSE loss values, whereas the rest are accuracies.}
\label{tab:traintestfinal}
\end{table}

\section{Experimental Results}

In this section we evaluate the capabilities of MIDI-VAE. Wherever mentioned, corresponding audio samples can be found on YouTube.
\footnote{\url{https://goo.gl/vb8Yrh}} 





\subsection{Style Transfer}\label{subsec:styletransfer}

To evaluate the effectiveness of MIDI-VAE's style transfer, we train three separate style evaluation classifiers. The input features are the pitch, velocity and instrument rolls respectively. The three style classifiers are also combined to output a voting based ensemble prediction. The accuracy of the classifiers is computed as the fraction of correctly predicted styles per bar in a song. We predict the likelihood of the source style \emph{before} and \emph{after} the style change. If the style transfer works, the predicted likelihood of the source style decreases. The larger the difference, the stronger the effect of the style transfer. Note that for all experiments presented in this paper we set the number of styles $k=2$, that is, one MIDI-VAE model is trained on two styles, e.g., Classic vs. Jazz. Therefore, the style classifier is binary and a reduction in probability of the source style is equivalent to an increase in probability of the target style of the same magnitude. All style classifiers use two-layer GRUs with a state size of 256.
Table~\ref{tab:styleensemble} shows the performance of MIDI-VAE's style transfer when measured by the ensemble style classifier. We trained a separate MIDI-VAE for each style pair. For each pair of styles we perform a style change on all songs in both directions and average the results. The style transfer works for all models, albeit to varying degrees. In all cases except for JvP, the predictor is even skewed below 0.5, meaning that the target style is now considered more likely than the source style. 

\begin{table}[t]
\tabcolsep=0.11cm
\centering
\begin{tabular}{|l?l|l|l?l|l|l|}
\hline
 & \multicolumn{3}{c?}{Train Songs} & \multicolumn{3}{c|}{Test Songs} \\ \hline
 & Before & After & Diff. & Before & After & Diff.\\ \Xhline{1pt}
CvJ & 0.92 & 0.38 & \textbf{0.54} & 0.87 & 0.39 & \textbf{0.48}\\ \hline
CvP& 0.94 & 0.43 & 0.51 & 0.92 & 0.45 & 0.47\\ \hline
JvP	& 0.72 & 0.60 & 0.12 & 0.72 & 0.62 & 0.10 \\ \hline
BvM& 0.77 & 0.45 & 0.32 & 0.66 & 0.47 & 0.19\\ \hline
\end{tabular}
\caption{Style transfer performance (ensemble classifier accuracies before and after) between all style pairs.}
\label{tab:styleensemble}
\end{table}

\begin{figure}[t]
\centering
\begin{tikzpicture}

\begin{axis}[
xlabel={Classic},
ylabel={Jazz},
xmin=-0.5, xmax=15.5,
ymin=-0.5, ymax=15.5,
width=0.8\columnwidth,
height=0.8\columnwidth,
xtick={0,1,2,3,4,5,6,7,8,9,10,11,12,13,14,15},
xticklabels={piano,percussion,organs,guitar,bass,strings,ensemble,brass,reed,pipe,synth lead,synth pad,synth effects,ethnic,percussive,sound effects},
ytick={0,1,2,3,4,5,6,7,8,9,10,11,12,13,14,15},
yticklabels={sound effects,percussive,ethnic,synth effects,synth effects,synth lead,pipe,reed,brass,ensemble,strings,bass,guitar,organs,percussion,piano},
tick align=outside,
yticklabel style = {font=\tiny},
xticklabel style = {rotate=90,font=\tiny},
tick pos=left,
x grid style={lightgray!92.02614379084967!black},
y grid style={lightgray!92.02614379084967!black},
colorbar,
colormap/viridis,
point meta min=0,
point meta max=1,
colorbar style={font=\tiny, ytick={0,0.2,0.4,0.6,0.8,1},yticklabels={0.0,0.2,0.4,0.6,0.8,1.0},ylabel={}}
]
\addplot graphics [includegraphics cmd=\pgfimage,xmin=-0.5, xmax=15.5, ymin=15.5, ymax=-0.5] {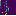};




\end{axis}

\end{tikzpicture}
\caption{The matrix visualizes how the instruments are changed when switching from Jazz to Classic, averaged over all Jazz songs in the test set.}
\label{fig:switchinstrmatrix}
\end{figure}

Table~\ref{tab:styleall} shows the style transfer results measured by each individual style classifier. We can see that pitch and velocity contribute equally to the style change, whereas instrumentation seems to correlate most with the style. For CvJ and CvP, switching the style heavily changes the instrumentation. Figure~\ref{fig:switchinstrmatrix} illustrates how the instruments of all songs in our Jazz test set are changed when switching the style to Classic. Only few instruments are rarely changed (piano, ensemble, reed), whereas most others are mapped to one or multiple different instruments. The instrument switch between genres with highly overlapping instrumentation (JvP, BvM) is much less pronounced. 
Classifying style based on the note pitches and velocities of one bar is more difficult, as shown by the \quotes{before} accuracies in Table~\ref{tab:styleall}, which are generally lower than the ones of the instrument roll based classifier. Nevertheless, the style transfer changes pitch and velocity towards the target style. 
MIDI-VAE retains most of the original melody, while often changing accompanying instruments to suit the target style. This is generally desirable, since we do not want to change the pitches so thoroughly that the original song cannot be recognized anymore.
We provide examples of style transfers on a range of songs from our training and test sets on YouTube (see \emph{Style transfer songs}). 

\begin{table}[tbh]
\centering
\begin{tabular}{|l?l|l?l|l?l|l|}
\hline
 & \multicolumn{2}{c?}{Pitch} & \multicolumn{2}{c?}{Velocity} & \multicolumn{2}{c|}{Instrument} \\ \hline 
 & Bf. & Af. & Bf. & Af. & Bf. & Af.\\ \Xhline{1pt}
CvJ Test 	& 0.77 & 0.66 & 0.67 & 0.57 & 0.90 & 0.20\\
\hline
CvP Test	& 0.77 & 0.67 & 0.71 & 0.60 & 0.91 & 0.27\\ \hline
JvP Test	& 0.65 & 0.63 & 0.67 & 0.64 & 0.67 & 0.55\\ \hline
BvM Test	& 0.55 & 0.47 & 0.60 & 0.49 & 0.64 & 0.47\\ \hline
\end{tabular}
\caption{Average before and after classifier accuracies for all classifiers (pitch/instrument/velocity) for the test set.}
\label{tab:styleall}
\end{table}

\subsection{Latent Space Evaluation}

\begin{figure}[t]
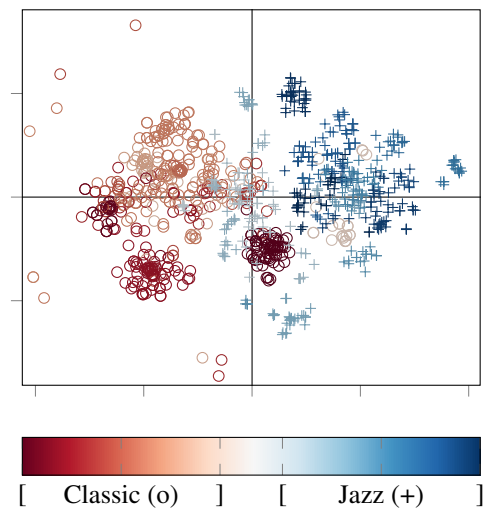

\centering
\include{figs/tsnescattersongs2}
\caption{t-SNE plot of latent vectors for bars from 20 Jazz and Classic songs. Bars from the same song were given the same color. Lighter colors mean that the ensemble style classifier was less certain in its prediction.}
\label{fig:tsnescatterlatentsongs}
\end{figure}

Figure~\ref{fig:tsnescatterlatentsongs} shows a t-SNE~\cite{maaten2008visualizing} plot of the latent vectors for all bars of 20 Jazz and 20 Classic pieces. The darker the color, the more \quotes{jazzy} or \quotes{classical} a song is according to the ensemble style classifier. The genres are well separated, and most songs have all their bars clustered closely together (likely thanks to the instrument roll being constant).
Some classical pieces are bleeding over into the Jazz region and vice versa. As can be seen from the light color, the ensemble style classifier did not confidently assign these pieces to either style. 


We further perform a sweep over all 256 latent dimensions on randomly sampled bars to check whether changing one dimension has a measurable effect on the generated music. We define 27 metrics, among which are total number of (held) notes, mean/max/min/range of (specific or all) pitches/velocities, and style changes. 
Besides the obvious dimensions where the style classifier is attached, we find that some dimensions correlate with the total number of notes played in a song, the highest pitch in a bar, or the occurrence of a specific pitch. The changes can be seen when plotting the pitches, but are difficult to hear. 
Furthermore, the dimensions are very entangled, and changing one dimension has multiple effects. Higher values for $\beta\in \{1,2,3\}$ slightly improve the disentanglement of latent dimensions, but strongly reduce reconstruction accuracy (see Figure~\ref{fig:betaepseval}). 
We added samples to YouTube to show the results of manipulating individual latent variables.



\subsection{Generation and Interpolation}

MIDI-VAE is capable of producing smooth interpolations between bars. This allows us to generate medleys by connecting short pieces from our dataset. 
The interpolated bars form a musically consistent bridge between the pieces, meaning that, e.g., pitch ranges and velocities increase when the target bar has higher pitch or velocity values.
We can also merge entire songs together by linearly interpolating the latent vectors for two bar progressions, producing interesting mixes that are surprisingly fun to listen to. The original songs can sometimes still be identified in the mixtures, and the resulting music sounds harmonic. 
We again uploaded several audio samples to YouTube (see \emph{Medleys}, \emph{Interpolations} and \emph{Mixtures}).




\section{Conclusion}
We introduce MIDI-VAE, a simple but effective model for performing style transfer between musical compositions. We show the effectiveness of our method on several different datasets and provide audio examples. Unlike most existing models, MIDI-VAE incorporates both the dynamics (velocity and note durations) and instrumentation of music. In the future we plan to integrate our method into a hierarchical model in order to capture style features over longer time scales and allow the generation of larger pieces of music. 
To facilitate future research on style transfer for symbolic music, and sequence tasks in general, we make our code and data publicly available.\footnote{\url{https://github.com/brunnergino/MIDI-VAE}}







\bibliography{ISMIRtemplate}

\end{document}